\documentclass[reprint,amsmath,amssymb,aip,floatfix]{revtex4-1}

\usepackage[utf8]{inputenc}
\usepackage[T1]{fontenc}
\usepackage{mathptmx}
\usepackage{etoolbox}
\usepackage{graphicx}
\usepackage{dcolumn}
\usepackage{bm}
\usepackage[version=4]{mhchem}
\usepackage{siunitx}
\usepackage{blkarray}
\usepackage{braket}
\DeclareSIUnit\angstrom{\text {Å}}
\DeclareSIUnit\atomicmassunit{u}
\DeclareSIUnit\elementarycharge{\text {\ensuremath {e}}}
\DeclareSIUnit\bohr{\text {\ensuremath {a}}_{0}}

\newcommand{\chead}[1]{\multicolumn{1}{c}{#1}}

\makeatletter
\def\@email#1#2{%
 \endgroup
 \patchcmd{\titleblock@produce}
  {\frontmatter@RRAPformat}
  {\frontmatter@RRAPformat{\produce@RRAP{*#1\href{mailto:#2}{#2}}}\frontmatter@RRAPformat}
  {}{}
}%
\makeatother
\begin{document}

\title[Equivariant machine learning of Electric Field Gradients -- Predicting the quadrupolar coupling constant in the \ce{MAPbI3} phase transition]{Equivariant machine learning of Electric Field Gradients -- Predicting the quadrupolar coupling constant in the \ce{MAPbI3} phase transition}

\author{Bernhard Schmiedmayer}
\email{bernhard.schmiedmayer@univie.ac.at}
\affiliation{%
University of Vienna, Faculty of Physics and Center for Computational Materials Science, Kolingasse 14-16, A-1090 Vienna, Austria
}

\author{J.W. Wolffs (Jop)}
\affiliation{%
Radboud University, Institute for Molecules and Materials, Heyendaalseweg 135, 6525 AJ Nijmegen, The Netherlands
}

\author{Gilles A. de Wijs}
\affiliation{%
Radboud University, Institute for Molecules and Materials, Heyendaalseweg 135, 6525 AJ Nijmegen, The Netherlands
}

\author{Arno P.M. Kentgens}
\affiliation{%
Radboud University, Institute for Molecules and Materials, Heyendaalseweg 135, 6525 AJ Nijmegen, The Netherlands
}

\author{Jonathan Lahnsteiner}
\affiliation{%
VASP Software GmbH, Berggasse 21/14, A-1090, Vienna, Austria
}

\author{Georg Kresse}
\affiliation{%
University of Vienna, Faculty of Physics and Center for Computational Materials Science, Kolingasse 14-16, A-1090 Vienna, Austria
}
\affiliation{%
VASP Software GmbH, Berggasse 21/14, A-1090, Vienna, Austria
}

\date{\today}

\begin{abstract}
    We present a strategy combining machine learning and first-principles calculations to achieve highly accurate nuclear quadrupolar coupling constant predictions. Our approach employs two distinct machine-learning frameworks: a machine-learned force field to generate molecular dynamics trajectories and a second model for electric field gradients that preserves rotational and translational symmetries. By incorporating thermostat-driven molecular dynamics sampling, we enable the prediction of quadrupolar coupling constants in highly disordered materials at finite temperatures. We validate our method by predicting the tetragonal-to-cubic phase transition temperature of the organic-inorganic halide perovskite \ce{MAPbI3}, obtaining results that closely match experimental data.
\end{abstract}

\maketitle
\section{Introduction}
\label{sec:intro}
Density functional theory (DFT) can compute specific nuclear magnetic resonance (NMR) properties, such as chemical shifts \cite{pickard2001all,yates2007calculation} and the quadrupolar coupling constant via the electric field gradient (EFG) \cite{petrilli1998electric}, but these calculations are usually based on a zero-temperature, ground-state framework. In contrast, NMR experiments are performed at finite temperatures, where dynamic effects ---such as vibrational, rotational, and conformational motions--- play a critical role in shaping the observed parameters. In systems like liquids, organic molecules, or dynamically disordered solids, these temperature-dependent dynamics can lead to significant variations in the measured NMR signals \cite{sattig2014nmr,dracinsky2016temperature}. Consequently, to obtain results that quantitatively match experimental data, it is essential to perform dynamic averaging over a statistically significant ensemble of configurations. This dynamic averaging often requires long simulation times and large system sizes (comprising hundreds of atoms) to accurately capture the full spectrum of molecular motions and to ensure the convergence of computed NMR parameters, particularly when slow dynamics such as molecular rotations are involved. Given the high computational cost of performing DFT-driven molecular dynamics ---especially when calculating properties like the EFG at frequent intervals--- machine learning (ML) techniques have emerged as a promising alternative for efficiently approximating these properties while maintaining the necessary level of accuracy \cite{behler2007generalized,balabin2009neural,bartok2010gaussian,bartok2018machine,jinnouchi2020fly,liu2022phase,charpentier2025first}.

\begin{figure}
    \centering    \includegraphics{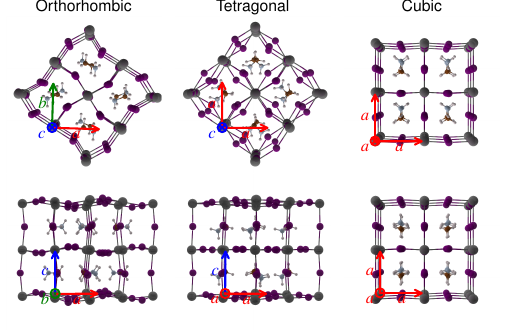}
    \caption{The phase sequence ---orthorhombic, tetragonal, and cubic--- upon heating of \ce{MAPbI3}. The arrows indicate the pseudocubic lattice vectors in the respective structure. Starting from a $2 \times 2 \times 2$ supercell with lattice vectors $\{\tilde{a}, \tilde{b}, \tilde{c}\}$, the pseudocubic lattice vectors $\{a, b, c\}$ are computed as $\{a,b,c\} = \{\sqrt{2}(\tilde{a}+\tilde{b})/4,\sqrt{2}(\tilde{a}-\tilde{b}))/4,\tilde{c}/2\}$. The orientations of the \ce{MA} molecules correspond to the predominant ordering patterns reported in the literature, albeit in the cubic phase, the \ce{MA} molecules become dynamically disordered \cite{jinnouchi2019phase}.}
    \label{fig:struc}
\end{figure}

The hybrid perovskite methylammonium lead iodide (\ce{MAPbI3}) serves as an excellent test case, as the slow rotational dynamics of the organic methylammonium (\ce{MA}) make straightforward first-principles (FP) molecular dynamics computationally expensive \cite{wasylishen1985cation,lahnsteiner2016room}. This material also possesses significant scientific and technological potential, as a promising solar cell material known for its high charge-carrier mobility \cite{hirasawa1994exciton,kojima2009organometal,samuel2013electron,cui2015recent}. \ce{MAPbI3} has been extensively studied both experimentally and theoretically \cite{weber1978CH3NH3PbX3,wasylishen1985cation,poglitsch1987dynamic,noriko1990calorimetri,kawamura2002structural,baikie2013synthesis,stoumpos2013semiconducting,filippetti2014hybrid,mattoni2015methylammonium,chen2015rotational,weller2015complete,lahnsteiner2016room,whitfield2016structures,bokdam2016role,franssen2017symmetry,bokdam2017assessing,lahnsteiner2018finite,schuck2018infrared,jinnouchi2019phase}. It has been shown that \ce{MAPbI3} undergoes two entropy-driven phase transitions: from an orthorhombic to a tetragonal phase at \SI{160}{\kelvin} and from the tetragonal to the cubic phase at \SI{330}{\kelvin}. In the orthorhombic phase, the MA is frozen in two specific orientations along the $a$ or $b$ axis, leading to significant distortions in the \ce{PbI6} octahedra. Upon transitioning to the tetragonal phase, the MA exhibits a two-dimensional disorder in the $ab$-plane, inducing axial symmetry in the octahedra, which rotate about the $c$-axis with a temperature-dependent angle \cite{weller2015complete}. As the material approaches the cubic phase, these distortions gradually diminish. At \SI{330}{\kelvin}, \ce{MAPbI3} transitions into a dynamically disordered state, where the \ce{MA} molecules undergo rapid reorientation, and the time-averaged structure exhibits cubic symmetry, despite the presence of instantaneous local distortions \cite{franssen2017symmetry,jinnouchi2019phase}. This phase transition results in a continuous reduction of octahedral distortions near the phase boundary. NMR studies on \ce{MAPbI3} have investigated the mobility of the MA, showing that its \ce{C}-\ce{N} axis is frozen in the orthorhombic phase, while increasing disorder characterises the tetragonal and cubic phases \cite{wasylishen1985cation,furukawa1989cationic,knop1990alkylammonium,xu1991molecular,baikie2015combined}. Further studies have demonstrated reorientation dynamics on the order of \SIrange{1.0}{5.4}{\pico\second}, along with interactions between the MA group and the lead-iodide octahedra, in the tetragonal and cubic phase \cite{poglitsch1987dynamic,noriko1990calorimetri,chen2015rotational,baikie2015combined}. Fig.~\ref{fig:struc} shows a structural model of the three phases of \ce{MAPbI3}.

\section{Method}
\label{sec:method}
\subsection{General remarks}
\label{sec:general}
In NMR spectroscopy of crystalline samples, the quadrupolar coupling constant $C_q$ is an experimentally measurable quantity, indicating the strength of the quadrupolar interaction of individual nuclei with the local EFG. It can be derived from the width of spectral patterns of powdered samples or from a combination of resonance frequencies in a nuclear quadrupole resonance spectrum. $C_q$ is given by
\begin{equation}
C_q = \frac{(eq)(eQ)}{h},
\end{equation}
where $e$ is the elementary charge, $eq$ is the largest principal component of the EFG, $Q$ is the nuclear quadrupole moment, and $h$ is Planck's constant \cite{kaupp2004calculation,autschbach2010analysis,muller2021solid}. For disordered materials such as \ce{MAPbI3}, it has been shown that averaging of the EFG tensor is essential to obtain $C_q$ values that closely match experimental results \cite{franssen2017symmetry,yamada2018static}. In this work, we focus on \ce{^{14}N}, a spin-1 nucleus, and \ce{^{127}I}, a spin-5/2 nucleus that exhibits a large quadrupolar coupling constant. As temperature increases, $C_q$ of \ce{^{14}N} decreases due to more effective cancellation of the EFG contributions from the four symmetry-related \ce{MA} orientations in the tetragonal phase, driven by changes in the $c/a$ ratio and MA orientation angles. In the cubic phase, where the \ce{MA} is dynamically disordered, rapid reorientations further average out any residual asymmetry, causing the largest principal component of the EFG to approach zero \cite{franssen2017symmetry}. Similarly, for \ce{^{127}I}, the quadrupolar coupling constants in different crystallographic layers converge towards a single finite value as their local environments become indistinguishable in the high-temperature cubic phase.

\subsection{Molecular dynamics simulations}
\label{sec:md}

In \ce{MAPbI3}, the reorientation of \ce{MA} occurs on the order of a few picoseconds -- significantly faster than the microsecond timescale observed in an NMR spectrum or even the nanosecond timescale of observed spin-relaxation \cite{wasylishen1985cation,poglitsch1987dynamic,onoda1990calorimetric,bakulin2015real}. As a result, during an NMR measurement, all possible \ce{MA} orientations within the \ce{PbI3} framework are effectively sampled. However, these experimentally relevant time scales are inaccessible to FP simulations, necessitating the use of surrogate models. In this study, we employ a machine-learned force field (MLFF) developed in our previous work \cite{schmiedmayer2024derivative}. The MLFF was trained on structures sampled from thermostat-driven MD simulations spanning \SIrange{80}{430}{\kelvin}. Initial training employed fixed-cell volumes, later refined using isothermal–isobaric simulations. The FP calculations were performed using the SCAN meta-GGA functional \cite{sun2015strongly}, which has been shown to yield accurate structural properties for \ce{MAPbI3} \cite{bokdam2017assessing}. All molecular dynamics (MD) calculations were performed using the machine learning framework integrated into the Vienna {\em Ab initio} Simulation Package (VASP) \cite{jinnouchi2019phase,jinnouchi2019fly,jinnouchi2020descriptors}.

The MD simulations were conducted in an isothermal-isobaric ($NpT$) ensemble with a time step of \SI{1}{\femto\second}. An $8 \times 8 \times 8$ supercell of \ce{MAPbI3} was used to ensure sufficient statistical sampling of the atomic dynamics. The system’s initial configurations and velocities were equilibrated over \num{100000} time steps using a Langevin thermostat with a friction coefficient of \SI{10}{\per\pico\second} to achieve the desired temperature \cite{allen2017computer,hoover1982high,evans1983computer}. The same thermostat settings were maintained during the subsequent production runs, which were carried out for \num{2000000} time steps, resulting in a total simulation time of \SI{2}{\nano\second}. To enable a time step of \SI{1}{\pico\second}, we increased the hydrogen mass by a factor of four. These trajectories were then used to compute the nuclear quadrupolar coupling constant $C_q$ and the $c/a$-ratio, with every \num{100}th step included in the averaging of the EFG tensor. 

\subsection{Machine learning model}
\label{sec:ml}
To characterise the atomic environment in our ML framework for the prediction of the EFG tensors, we utilise the $\lambda$-SOAP (Smooth Overlap of Atomic Potentials) descriptors developed by Grisafi {\em et al.} \cite{bartok2013representing,grisafi2018symmetry}. The core idea is to perform a series expansion of the smoothed atomic density $\rho_i(\textbf{r})$ around a central atom $i$, using spherical harmonics $Y_{l,m}(\textbf{r})$, and radial functions $\chi_{nl}(r)$:
\begin{equation}\label{eq:density}
    \rho_i(\textbf{r}) = \sum_j g_\eta(\textbf{r}-\textbf{r}_{ij}) f_\text{cut}(r_{ij}) = \sum_{nlm} c^i_{nlm} Y_{l,m}(\textbf{r}) \chi_{nl}(r).
\end{equation}
Here $\textbf{r}_{ij}$ is the vector from the central atom $i$ to a neighbouring atom $j$, and $r$ denotes the radial distance. The function $g_\eta(\textbf{r})$ is a three-dimensional Gaussian of width $\eta$, and $f_\text{cut}(r)$ is a Behler-Parrinello cut-off function \cite{behler2007generalized} designed to avoid abrupt discontinuities beyond a specified cut-off radius $R_{\text{cut}}$. The expansion coefficients $c^i_{nlm}$ are used to construct the descriptor of the atomic environment of atom $i$. For practical computation, the indices are truncated such that $n \in \mathbb{Z} \cap [1,N_{\text{max}}]$ and $l \in \mathbb{Z} \cap [0,L_{\text{max}}]$, while $m$ runs over the range $m \in \mathbb{Z} \cap [-l,l]$. $N_{\text{max}}$ and $L_{\text{max}}$ represent the radial and angular resolution of the descriptor, respectively. Along with $\eta$ and $R_{\text{cut}}$, they serve as hyperparameters of the ML framework that require optimisation. As radial functions, we used normalised spherical Bessel functions of the first kind $\chi_{nl}(r) = \hat{j}_l(q_{nl} r)$, where the parameters $q_{nl}$ are chosen such that $q_{nl}R_\text{cut}$ is the $l$th root of $j_n$. To distinguish between atomic species, we extend the expansion coefficients to $c^{iJ}_{nlm}$ by summing only over neighbours belonging to a specific atomic species $J$:
\begin{equation}\label{eq:expansion}
    c^{iJ}_{nlm} = \sum_{j \in J} h_{nl}(r_{ij}) Y_{lm}(\textbf{r}_{ij}),
\end{equation}
where $h_{nl}(r)$ is the radial projection kernel to the expansion coefficient. It combines Gaussian smoothing of the atomic density with a cut-off function, and projects the smoothed and truncated density onto the radial basis functions $\chi_{nl}(r)$.

The EFG tensor, $\textbf{V}$, is a traceless second-rank tensor \cite{kaupp2004calculation}. As a symmetric tensor, it generally consists of six independent components; however, the traceless condition imposes a constraint that reduces the number of independent components to five. One way to achieve this reduction is to express the EFG in its irreducible spherical tensor (IST) representation of rank $l=2$. In constructing our machine learning framework, it is essential to respect the symmetry properties of the EFG when describing the local atomic environments. The transformation properties of the spherical harmonics ---as used in the calculation of $c^{iJ}_{nlm}$ in Eq.~\eqref{eq:expansion}--- for $Y_{l=2,m}$ are identical to those of the IST representation of the EFG along $m$. When considering only pairwise interactions, constructing a two-body descriptor that preserves these transformation properties is straightforward; it suffices to select only expansion coefficients where $l=2$, such as $^m\textbf{D}^{(2)}_i = c^{iJ}_{n,l=2,m}$. Here, $J$ and $n$ denote the dimensions of the feature space for the descriptor. The expansion coefficients can be combined with Clebsch–Gordan coefficients $\braket{\lambda\mu|l'm'lm}$ to capture higher-order interactions. In this context, $\lambda=2$ is the rank of the tensor, and $\mu$ indexes its IST components. This approach extends naturally to three-body descriptors, expressed as follows \cite{grisafi2018symmetry}:
\begin{equation}
^\mu\textbf{D}^{(3)}_i = \sum_{mm'}c^{iJ}_{nlm}c^{iJ'}_{n'l'm'} \braket{2\mu|l'm'lm},
\end{equation}
where $n$, $n'$, $l$, $l'$, $J$, and $J'$ define the feature space of $\textbf{D}$. This formulation allows the descriptor to capture three-body correlation effects while maintaining consistency with the symmetry properties of the EFG tensor.

The EFG in its IST representation $\tilde{\textbf{V}}_i$, at a given atom $i$, can be expressed in terms of its atomic environment and, by extension, its descriptor $\textbf{D}^\mu_i$ as:
\begin{equation}
\tilde{V}^\mu_i = \sum_{I_{\text{ref}},\nu} \omega^\nu_{I_\text{ref}} \textbf{D}^\mu_i \cdot \textbf{D}^\nu_{I_\text{ref}}.
\end{equation}
Here, $\mu$ indexes the IST components of the EFG tensor, while the regression weights $\omega^\nu_{I_\text{ref}}$ serve as fitting parameters. For each reference environment $I_{\text{ref}}$, the model evaluates the dot product $\mathbf{D}^\mu_i \cdot \mathbf{D}^\nu_{I_\text{ref}}$ to quantify the similarity between the environment of atom $i$ and that reference. This similarity is then weighted by $\omega^\nu_{I_\text{ref}}$. The set of all reference descriptors $\{\textbf{D}^\nu_{I_\text{ref}}\}$ serves as the kernel basis for the regression. The weights are determined using a linear regression via the least squares method, allowing simultaneous optimisation across all training configurations and EFG tensor components. Rather than using every descriptor from the training set, we employ sparse regression to choose a smaller subset of reference descriptors as our kernel basis. The number of reference descriptors is directly related to the number of fitting parameters $M$. In this context, $M$ is an additional hyperparameter of the model. Choosing too few reference descriptors leads to underfitting and poor accuracy, while selecting too many increases the risk of overfitting, ultimately degrading predictive performance. The selection of reference descriptors is performed using farthest-point sampling. This method for learning EFG tensors closely parallels the work of T. Charpentier \cite{charpentier2025first}. Further work on machine learning of EFGs can be found in Refs.~\onlinecite{harper2024performance,shakiba2024machine,mahmoud2025graph,sun2025machine}.

We constructed three datasets of EFGs: two for training and one for validation. The configurations were selected from \ce{NpT} MD trajectories at \SI{225}{K}, \SI{275}{K}, and \SI{325}{K} generated using our MLFF \cite{schmiedmayer2024derivative}.  Each dataset consists of an equal number of configurations per temperature. For the training datasets, configurations were selected using farthest-point sampling, ensuring diversity in atomic environments. This was achieved by first computing the descriptor for each nitrogen atom in the dataset and then iteratively selecting configurations that contain the most unique atomic environments. In contrast, configurations for the validation dataset were chosen randomly. One training set and the validation set were generated from a $4\times4\times4$ supercell, the other training set was based on a $2\times2\times2$ supercell. Due to the difference in system sizes, the total number of configurations varies: \num{150} for the $4\times4\times4$ training set and \num{1200} for the $2\times2\times2$ training set. However, both training datasets contain the same number of data points per atomic species: \ce{Pb} -- \num{9600}, \ce{I} -- \num{28800}, \ce{C} -- \num{9600}, \ce{N} -- \num{9600}, and \ce{H} -- \num{57600}. The validation set comprises \num{30} configurations.

The EFG calculations for the datasets were performed using the method of Petrilli {\em et al.} \cite{petrilli1998electric}, within VASP \cite{kresse1993ab,kresse1996efficiency,kresse1996efficient} using the projector-augmented wave (PAW) \cite{blochl1994projector,kresse1999ultrasoft} method and the Perdew-Burke-Ernzerhof (PBE) exchange-correlation functional \cite{perdew1996generalized}. To ensure high precision, a plane-wave energy cut-off of \SI{800}{\electronvolt} was employed. Hard pseudopotentials were used for all atomic species of \ce{MA}. More details on the potentials used can be found in Tab.~\ref{tab:pot}. The Blocked-Davidson algorithm was applied for electronic minimisation, with an electronic convergence criterion of \SI{1e-6}{\electronvolt} and Gaussian smearing of \SI{0.05}{\electronvolt}. A minimum of \num{10} self-consistent electronic iterations was enforced. The use of symmetry was disabled. A single $\Gamma$-point was used for the $4\times4\times4$ supercell, while a $2\times2\times2$ Monkhorst-Pack mesh was employed for the $2\times2\times2$ supercell. 

\begin{table}[htb]
    \centering
    \caption{%
    PAW potentials used for the EFG calculations in \ce{MAPbI3}.
    }
    \label{tab:pot}
    \begin{tabular}{llccc}
        \hline
        \hline
        \chead{Atom} & \chead{Label} & valence orbitals & $r$ cut-off (\si{\bohr}) \\
        \hline
        \ce{H}  & \verb|PAW_PBE H_h 06Feb2004| & 1s\phantom{2p}   & 0.7 \\
        \ce{C}  & \verb|PAW_PBE C_h 20Jan2021| & 2s2p & 1.1 \\
        \ce{N}  & \verb|PAW_PBE N_h 19Jan2021| & 2s2p & 1.1\\ 
        \ce{I}  & \verb|PAW_PBE I 08Apr2002|   & 5s5p & 2.3\\
        \ce{Pb} & \verb|PAW_PBE Pb 08Apr2002|  & 6s6p & 3.1\\
        \hline
        \hline
    \end{tabular}
\end{table}

\begin{figure}[htb]
    \centering
    \includegraphics{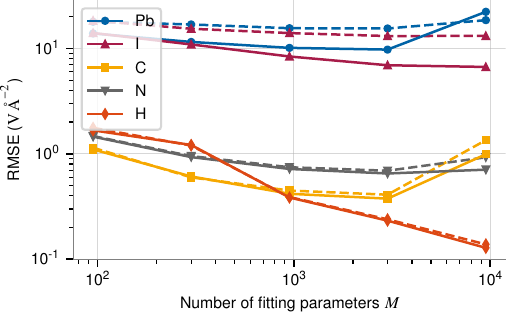}
    \caption{%
    Error on a validation data set of \num{30} $4\times 4 \times 4$ structures evenly distributed over three temperatures; \SI{225}{\kelvin}, \SI{275}{\kelvin}, and \SI{325}{\kelvin}. The dashed line is trained on a $2\times 2 \times 2$ supercell, while the solid line uses a $4\times 4 \times 4$ supercell. Both training data sets consist of the same number of data points per atomic species. Optimal fitting parameters are specified in the main text.
    }
    \label{fig:scan}
\end{figure}

Only three-body descriptors were used in the model. We initially explored two-body descriptors as well, but found that the inclusion of two-body descriptors did not lead to any significant improvement in the fit. All hyperparameters ($N_{\text{max}}$ -- radial resolution, $L_{\text{max}}$ -- angular resolution, $\eta$ -- Gaussian broadening, $R_{\text{cut}}$ -- cut-off radius, $M$ -- number of fitting parameters) were optimised via a grid search. The optimal parameters for nitrogen were determined as $N_{\text{max}} = \num{8}$, $L_{\text{max}} = \num{4}$, $\eta = \SI{0.4}{\angstrom}$, and $R_{\text{cut}} = \SI{8.5}{\angstrom}$. Since $2R_{\text{cut}}$ exceeds the system sizes in the $2 \times 2 \times 2$ dataset, we examined the errors on the validation set when training on different system sizes. A learning curve illustrating the dependence of the error on the number of fitting parameters, $M$, and the different training sets is shown in Fig.~\ref{fig:scan}.

\begin{figure*}[htb]
    \centering
    \includegraphics{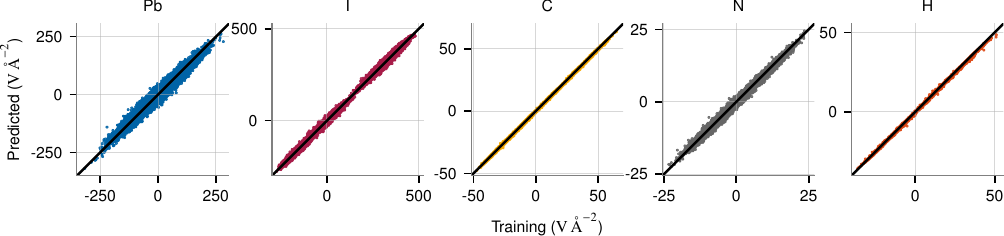}
    \caption{%
    Scatter plot of the training fit for all Cartesian components of the EFG across all atomic species of \ce{MAPbI3}.
    }
    \label{fig:scatter}
\end{figure*}

For the final fit, we used the training set consisting of $4\times4\times4$ structures only, as it resulted in lower errors across all species. The errors were evaluated on the test set using the Frobenius norm, normalised by the number of matrix elements (nine), which corresponds to the root mean square error (RMSE) of all matrix elements. Since the RMSE is directly proportional to the magnitude of the property, and the quadrupolar coupling varies significantly between atomic species, a direct comparison across species can be misleading. To provide a more balanced assessment, we scale the RMSE by the inverse of the standard deviation $\sigma^{-1}$ of the EFGs from the training set for each species. The scaled RMSE is presented in addition to the unscaled RMSE to give a more comprehensive evaluation of the model performance in Tab.~\ref{tab:err}. A scatter plot of the training fit is shown in Fig.~\ref{fig:scatter}.

\begin{table}[htb]
    \centering
    \caption{%
    Number of fitting parameters $M$ per atomic species and the resulting RMSE of the fit. The relative error, defined as the RMSE normalised by the standard deviation $\sigma$ of the training set, is also provided.
    }
    \label{tab:err} 
    \begin{tabular}{lccS}
        \hline
        \hline
        \chead{Atom} & \chead{$M$} & \chead{RMSE (\si{\volt\per\square\angstrom})} & \chead{$\text{RMSE}/\sigma$ (\si{\percent})} \\
        \hline
        \ce{H}  & 9500 & 0.14 &  1.2 \\
        \ce{C}  & 3000 & 0.37 &  1.9 \\
        \ce{N}  & 3000 & 0.63 &  9.9 \\
        \ce{I}  & 9500 & 6.36 &  4.8 \\
        \ce{Pb} & 3000 & 9.36 & 21.1 \\
        \hline
        \hline
    \end{tabular}
\end{table}

It is important to note that errors for atomic species other than nitrogen could be further reduced by choosing different hyperparameters (not shown). This is particularly true for the \ce{PbI3} framework ions, where increasing the cut-off radius lowers the error when additional \ce{PbI3} environments are included. This suggests that long-range interactions are more important for some species. In contrast, hydrogen exhibits the lowest error with a shorter cut-off radius. Furthermore, increasing $N_{\text{max}}$, {\em i.e.}, the radial resolution, along with a larger $R_{\text{cut}}$ improves the fit. However, this comes at a steep computational cost: The number of descriptor components scales as $N_{\text{max}}^2$, while a larger $R_{\text{cut}}$ increases the number of neighbouring atoms contributing to the descriptor. Beyond a certain point, the improvement in accuracy becomes marginal and does not justify the additional cost. For nitrogen, we selected what we consider to be the best fit by balancing accuracy and computational efficiency. Additionally, Fig.~\ref{fig:scan} shows that the errors for \ce{I} and \ce{H} have not yet reached their minimum with the current number of regression parameters. Both species have significantly more data points in the training set than the other elements ---three times more for \ce{I} and six times more for \ce{H} compared to \ce{Pb}, \ce{C}, and \ce{N}--- indicating that increasing the number of fitting parameters could further reduce the error for \ce{Pb}, \ce{C}, and \ce{N} if more training data points were included. 

Additionally, we adopted the density-based long-range electrostatic descriptors developed by Faller \textit{et al.} \cite{faller2024density} for use with tensorial quantities, such as the EFG. However, these descriptors were ultimately not included in our final model, as they did not improve the overall EFG prediction accuracy. The only exception was lead, where incorporating long-range descriptors reduced the error compared to the $\lambda$-SOAP-based model. For lead, the absolute error decreased to \SI{8.3}{\volt\per\square\angstrom}, with a corresponding normalised error ---defined as the error divided by the standard deviation--- of \SI{19}{\percent}. The $\lambda$-SOAP errors for lead are reported in Table~\ref{tab:err}. The minor decrease in error indicates that the residual errors are only marginally related to the model's finite cut-off, but rather to the neglect of higher body-order interactions and the restriction to linear regression. For the present purpose ---the temperature-dependent description of the nuclear quadrupolar coupling constant--- a \SIrange{5}{10}{\percent} relative accuracy is acceptable (see Sec.~\ref{sec:results}).

\section{Results and Discussion}
\label{sec:results}

\ce{MAPbI3} undergoes two phase transitions upon heating: from an orthorhombic to a tetragonal phase, and subsequently from the tetragonal to a cubic phase. We estimated the orthorhombic–tetragonal transition temperature by performing a temperature-ramped MD simulation under thermostat control and monitoring the evolution of the pseudocubic lattice parameters. This approach yielded a transition temperature of approximately \SI{170}{\kelvin}. Since the simulation proceeds in finite temperature steps, the system can undergo superheating. This tends to shift the apparent transition point of our model to higher temperatures and could lead to a systematic overestimation. The experimental orthorhombic-tetragonal phase transition is reported at \SI{165}{\kelvin} \cite{weller2015complete}.

\begin{figure*}[htb]
    \centering
    \includegraphics{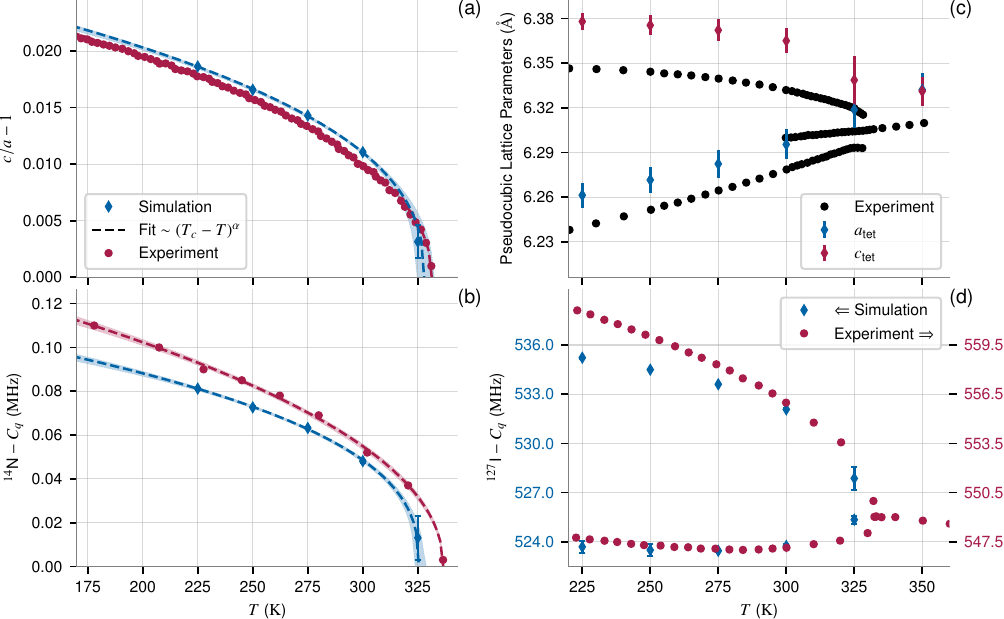}
    \includegraphics{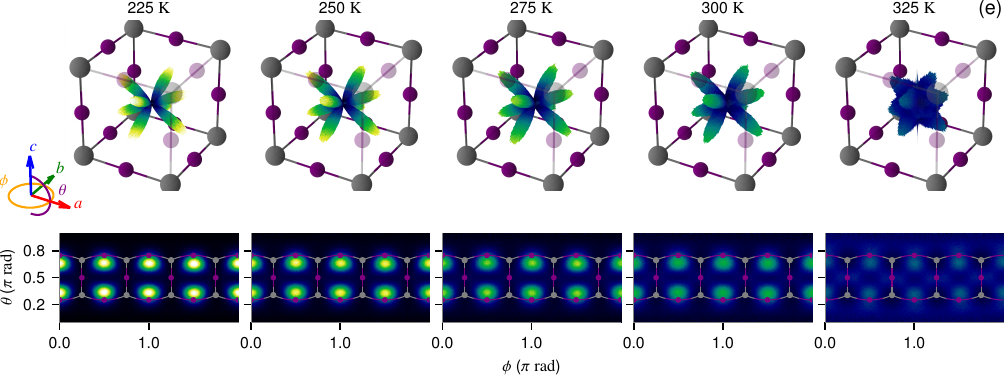}
    \caption{%
    \textbf{(a)} The simulated $c/a$ ratio of the simulation box alongside experimental data from Weller {\em et al.} \cite{weller2015complete}. The dashed lines represent fits using a power law of the form $A(T_c - T)^\alpha$. Error bars for the simulated results were computed using block averaging. The statistical uncertainties of the fit are indicated by the shaded regions, which represent a $2\sigma$ confidence interval.  These were obtained by generating \num{10000} random curves based on a multivariate normal distribution centred around the actual fitting parameters, and distributed according to the covariance matrix of the fit. The fit was performed using the Levenberg-Marquardt algorithm \cite{more1977levenberg} as implemented in \texttt{scipy.optimise.curve\_fit}\cite{2020SciPy-NMeth}. \textbf{(b)} Temperature dependence of the quadrupolar coupling constant $C_q$ (in \si{\mega\hertz}) for \ce{^{14}N} from both experiment and simulation. The experimental values were obtained via powder \ce{^{14}N} NMR and are taken from Ref.~\onlinecite{franssen2017symmetry}. \textbf{(c)} Pseudocubic lattice parameters extracted from the MD trajectories, shown alongside experimental values from Ref.~\onlinecite{whitfield2016structures}. Error bars represent the standard deviation across the simulation data. \textbf{(d)} Temperature dependence of the quadrupolar coupling constant $C_q$ (in \si{\mega\hertz}) for \ce{^{127}I}, comparing experimental and simulated results. Assuming an asymmetry parameter $\eta$ of approximately zero for perovskites, the experimental data from Ref.~\onlinecite{yamada2018static} have been scaled by the nuclear spin prefactor of $2I(2I-1)/3$ for $I=5/2$.  Error bars on the simulation results were estimated using block averaging. A secondary $y$-axis is used in the plot: the left $y$-axis corresponds to the simulation results, while the right $y$-axis displays the experimental values. This approach allows for direct visual comparison, as the simulated results are systematically underestimated by approximately \SI{4.5}{\percent}. \textbf{(e)} Orientation distribution of the \ce{MA} molecules at various temperatures. A 2D heatmap and corresponding 3D polar plots visualise the angular distributions. A schematic \ce{PbI3} cage is overlaid for reference. The \ce{MA} molecules are roughly coplanar with four neighbouring iodine atoms. Above and below the \ce{MAI} plane, a \ce{PbI2} plane is located, and the hydrogen atoms in the \ce{NH3} group typically attach to one of the four iodine atoms above or below in the \ce{PbI2} plane. This is particularly clear in the \SI{225}{\kelvin} panel with bright spots close to the iodine atoms ($\theta$ around \SI{0.3}{\pi} and \SI{0.6}{\pi}).
    }
    \label{fig:cq}
\end{figure*}

The tetragonal-cubic transition temperature was estimated between \SIrange{331}{335}{\kelvin} by analysing the critical behaviour of both $C_q$ of \ce{^{14}N} and the $c/a$ ratio of the pseudocubic lattice vectors. The temperature dependence of these quantities, alongside experimental reference data, is presented in Fig.~\ref{fig:cq}a,b. Near the phase transition, both $C_q$ and the $c/a$ ratio were fitted with a power-law function of the form $A(T_c - T)^\alpha$, where $T_c$ is the critical/transition temperature and $\alpha$ is the critical exponent. In principle, $T_c$ and $\alpha$ should be consistent across both observables, as they reflect the same underlying phase transition. For our simulated results, error bars were determined via block averaging, assuming an exponentially decaying autocorrelation function \cite{box2015time}. Details of the EFG tensor averaging methodology are discussed in Sec.~\ref{sec:EFGdyn}.

The optimised fitting parameters and their uncertainties are listed in Tab.~\ref{tab:fit}, alongside corresponding fits to the experimental data for comparison. The tetragonal-cubic transition temperature obtained from our simulations agrees with experimental values to within \SI{2}{\percent}. Although even closer agreement has been achieved using MLFFs for other materials ---see, for example, \cite{liu2022phase}--- our results are particularly strong given the complexity of \ce{MAPbI3}. A previous SCAN-based MLFF study using VASP employed a similar approach to estimate the transition temperature $T_c$ of \ce{MAPbI3}\cite{jinnouchi2019phase}. However, the reported value of \SI{353}{\kelvin} overestimated the experimental result by \SI{6.6}{\percent}. Notably, the simulations involved significantly fewer MD steps and smaller simulation boxes, which may have impacted accuracy. Additionally, the present MLFF was trained on a more refined dataset, likely enhancing its predictive power. Given these considerations, it is plausible that our results offer a more "accurate" description of the phase transition behaviour in \ce{MAPbI3}. The fitted critical exponents $\alpha$ differ between experiment and simulation, with the simulated transition appearing slightly more gradual. Nonetheless, the qualitative behaviour is well captured. Moreover, the consistent critical behaviour observed across both the $c/a$ ratio and $C_q$ reinforces their reliability as order parameters for describing the tetragonal-cubic phase transition.

\begin{table}[htb]
    \caption{%
    Fitting parameters and error estimates for the power law $A(T_c - T)^\alpha$ applied to the $c/a$ ratio and $C_q$ in both experiment and simulation. Here, $T_c$ is the phase transition temperature to the cubic phase, and $\alpha$ is the critical exponent. In principle, $T_c$ and $\alpha$ should be consistent across all fits.
    }
    \label{tab:fit}
    \sisetup{separate-uncertainty=true}
    \begin{tabular}{c|S[table-format=3.2(2)]S[table-format=1.3(2)]|S[table-format=3.1(2)]S[table-format=1.2(2)]}
        \hline
        \hline
               & \multicolumn{2}{c|}{Experiment} & \multicolumn{2}{c}{Simulation} \\
               & \chead{$T_c$ (\si{\kelvin})} & \multicolumn{1}{c|}{$\alpha$} & \chead{$T_c$ (\si{\kelvin})} & \chead{$\alpha$} \\
        \hline
        $C_q$  & 335.2\pm3.2 & 0.46\pm0.04 & 325.5\pm1.3 & 0.37\pm0.02 \\
        $c/a$  & 331.30\pm0.06 & 0.466\pm0.002 & 327.9\pm2.1 & 0.40\pm0.02 \\
        \hline
        \hline
\end{tabular}
\end{table}

In addition to reporting the quadrupolar coupling constants $C_q$ for \ce{^{14}N}, we also present values for \ce{^{127}I} in Fig.~\ref{fig:cq}d. Two distinct $C_q$ values are observed, corresponding to iodine atoms in different perovskite layers along the $c$-axis of the tetragonal \ce{MAPbI3}: the \ce{MAI}-layer and the \ce{PbI2}-layer. Iodine atoms in the \ce{MAI}-layer exhibit a higher $C_q$.  This can be attributed to their nearly linear \ce{Pb-I-Pb} bond angle, a geometry that maximises the electric field gradient. As the temperature increases and the material approaches the cubic phase, the local environments of iodine atoms in both layers become more similar, leading to the convergence of the two $C_q$ values. This trend is consistent with the temperature evolution of the pseudocubic lattice parameters shown in Fig.~\ref{fig:cq}e. Moreover, while in the tetragonal phase the \ce{MA} molecules preferentially hydrogen‑bond to iodine atoms in the \ce{PbI2}-layer, the orientation distribution in Fig.~\ref{fig:cq}e shows that at higher temperatures \ce{MA} increasingly samples all iodines of the \ce{PbI3} cage, reflecting a more uniform occupation of binding sites. This further supports the increasing equivalence of the iodine environments, explaining the convergence observed in the two $C_q$ values. Quantitatively, the simulated $C_q$ values are approximately \SI{4.5}{\percent} lower than the experimental values, while there is a slight overestimation of the perovskite volume.

\begin{figure}[htb]
    \centering
    \includegraphics{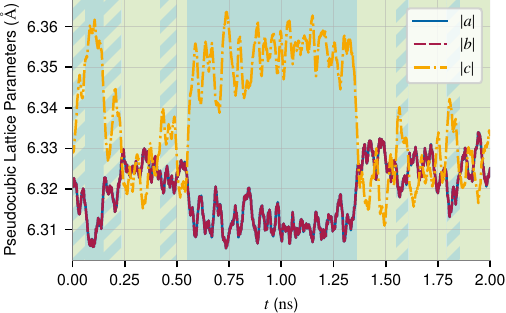}
    \caption{%
    Time evolution of the pseudocubic lattice parameters at \SI{325}{\kelvin}, every \num{100}th time step and smoothed using a running average over \num{150} snapshots. Critical fluctuations can be observed. The different phases are highlighted with different background colours.
    }
    \label{fig:MD}
\end{figure}

It is important to note that at \SI{325}{\kelvin}, both the $c/a$ ratio and $C_q$ show larger statistical error bars. During the MD trajectory at this temperature, multiple critical fluctuations between the tetragonal and cubic phases were observed, see Fig.~\ref{fig:MD}. After careful analysis, we decided to average the EFG over these fluctuations. To validate this approach, we considered three scenarios: (1) extracting time windows where the cubic phase was present, (2) using only tetragonal-phase configurations, and (3) averaging over the full MD trajectory. Experimental observations suggest that the tetragonal-to-cubic phase transition is tricritical, which means that both phases coexist within a certain temperature range \cite{whitfield2016structures}. However, determining the exact ratio of the two phases at a given temperature is challenging. In actual NMR experiments, a mixture of both phases would likely be observed. Thus, computing the time-averaged EFG over both phases provides the most consistent approach. Treating the last data point as purely tetragonal results in transition temperatures closer to the experiment, but worsens the overall fit. In contrast, treating it as cubic requires excluding it from the fit, significantly reducing accuracy. Given the evidence for phase coexistence and the improved fit quality, we conclude that averaging over the phase fluctuations is the most appropriate approach.

\section{Reorientation dynamics and EFG averaging}\label{sec:EFGdyn}

\begin{figure}[htb]
    \centering
    \includegraphics{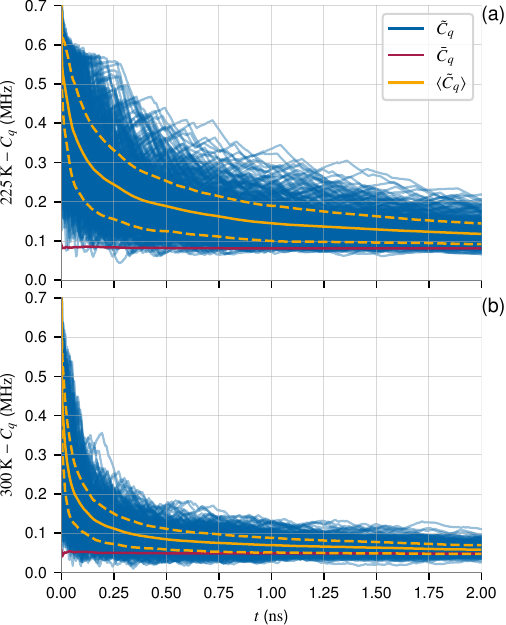}
    \includegraphics{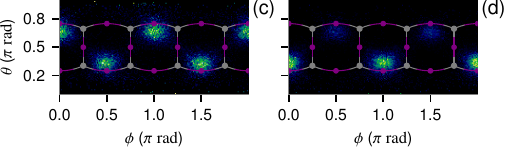}
    \caption{%
    \textbf{(a)} Time evolution of the quadrupolar coupling constant $C_q$ of nitrogen at \SI{225}{\kelvin}, computed using time-averaged electric field gradients. The simulation includes \num{512} nitrogen atoms. Blue lines represent the time-averaged quadrupolar coupling $\tilde{C}_q$ for individual atoms, while the yellow solid line shows the mean time-averaged coupling $\langle \tilde{C}_q \rangle$, with dashed yellow lines indicating its standard deviation. The red line corresponds to the quadrupolar coupling $\bar{C}_q$ obtained from the globally averaged electric field gradient. While $\tilde{C}_q$ and $\langle \tilde{C}_q \rangle$ converge gradually, $\bar{C}_q$ stabilizes significantly faster. \textbf{(b)} Time evolution of the quadrupolar coupling constant $C_q$ of nitrogen at \SI{300}{\kelvin}. \textbf{(c)} Orientation distribution of a single \ce{MA} molecule with an low individual $C_q$ value. All nearby iodine bonding sides are occupied with roughly equal probability, indicating uniform sampling. \textbf{(d)} Orientation distribution of a single \ce{MA} molecule with an relatively high individual $C_q$ value. Certain bonding sites are visited less frequently, due to steric hindrance from the \ce{MA} molecule positioned above, reflecting anisotropic sampling caused by corner-sharing constraints.
    }
    \label{fig:average}
\end{figure}

In the high-temperature phases of \ce{MAPbI3}, the \ce{MA} molecules exhibit both reorientation in the $ab$-plane and the $c$-direction within the \ce{PbI3} framework. The \ce{NH3} group of each \ce{MA} molecule forms weak hydrogen bonds with one of the iodine atoms of the surrounding cage. In the tetragonal phase, octahedral tilting brings two iodine atoms in each \ce{PbI2}-layer closer to the encapsulated \ce{MA} molecule, while the other two iodine atoms are offset toward neighbouring \ce{MA} molecules. This geometry produces four favourable hydrogen-bonding sites for each molecule -- two above and two below the \ce{MA}. Consequently, there are two preferred orientation families, depending on the local environment of the \ce{MA} molecule. These two families are related by a \SI{90}{\degree} rotation in the $ab$-plane and arrange themselves in a chequerboard pattern throughout the crystal. Fig.~\ref{fig:cq}e shows a combination of the two orientation distribution families, where each lobe corresponds to one hydrogen atom of the \ce{NH3} group bonded to one of the iodine atoms in the \ce{PbI2}-layers. For comparison, Fig.~\ref{fig:average}c is the orientation distribution of a single \ce{MA} molecule, {\em i.e.} one of the two families.

By symmetry, the two \ce{MA} orientation families produce the same time-averaged EFG tensor at the nitrogen site. Within each distinct family, the \ce{MA} molecules undergo rapid reorientations that possess an \SI{180}{\degree} rotational symmetry within the $ab$-plane. This dynamic symmetry strongly constrains the EFG tensor by averaging out its components in the $ab$-plane over time. As a result, the components in the $c$-direction become the dominant contributor. Since the local inorganic environment is similar across both families, the resulting time-averaged EFG tensors are equivalent regardless of the specific orientation family.

\begin{figure}[htb]
    \centering
    \includegraphics{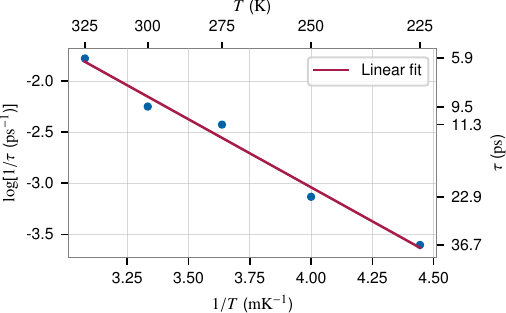}
    \caption{Arrhenius linear plot of the \ce{MA} reorientation times, extracted from the Laplace-transformed autocorrelation function at various temperatures. The linear fit is shown as a solid line, with its slope corresponding to the energy barrier of \SI{115\pm9}{\milli\electronvolt} for \ce{MA} reorientations.}
    \label{fig:arrhenius}
\end{figure}

Each \ce{MA} molecule should, on average, adopt all four orientations of its family equally. However, this does not occur on a \SI{2}{\nano\second} timescale for all \ce{MA} molecules. Intriguingly, all \ce{MA} molecules still display rapid reorientation motions on the timescale of \SI{35}{\pico\second} at \SI{225}{\kelvin}, accelerating to around \SI{5.5}{\pico\second} at \SI{325}{\kelvin}, see Fig.~\ref{fig:arrhenius}. These values were determined by performing the Laplace transform of the auto-correlation function of the \ce{MA} molecule orientations ---the direction of the \ce{C-N} bond--- over the \SI{2}{\nano\second} trajectory and are consistent with previously reported values in the literature \cite{poglitsch1987dynamic,noriko1990calorimetri,chen2015rotational,baikie2015combined,jinnouchi2019phase}. The slow reorientation is evident when the EFG tensor for each nitrogen atom is cumulatively averaged over the \SI{2}{\nano\second} trajectory, then diagonalised to yield a $C_q$ for each nitrogen atom, and finally, this scalar value is averaged over all nitrogen atoms (see the yellow line in Fig.~\ref{fig:average}a). We observe a slow decay that does not fully converge on the \SI{2}{\nano\second} timescale. However, averaging the EFG tensor over all nitrogen atoms before diagonalisation yields not only much smaller values for $C_q$ but also converges much more rapidly with respect to the total simulation time.

This result warrants closer examination. While the \ce{MA} molecules reorient frequently, their full rotational freedom is hindered by local interactions. Fig.~\ref{fig:average}c,d illustrate the orientation distribution for two specific \ce{MA} molecules during our \SI{2}{\nano\second} simulation at \SI{225}{\kelvin}. In Fig.~\ref{fig:average}c, the molecule spends equal amounts of time bonded to iodine atoms in the \ce{PbI2}-layer above and below, leading to a balanced orientation of the nitrogen atom. This results in a comparatively small $C_q$ value, approaching the value obtained by time-averaging the EFG tensor over all nitrogen atoms before diagonalisation. In contrast, Fig.~\ref{fig:average}d shows a molecule whose \ce{NH3} group spends the majority of the simulation time bonded to iodine atoms in the lower \ce{PbI2}-layer. While partial reorientation still occurs ---evident by the two lobes in the orientation distribution--- the occupation time of the upper bonding sites is minimal. This imbalance leads to a comparatively large value for $C_q$ of \SI{0.22}{\mega\hertz}.

The hindered motions originate from hydrogen-bonding constraints. Each iodine atom can accept only one hydrogen bond from an \ce{NH3} group. As a result, certain iodine sites may become blocked by neighbouring \ce{MA} molecules. These interactions induce correlation effects ---particularly along the $c$-direction--- that suppress free reorientation. The concerted reorientation processes are kinetically slow, developing over several nanoseconds, as indicated by the very slow decay of the solid yellow line in Fig.~\ref{fig:average}a. This slow decay indicates incomplete decorrelation within the available simulation window. At higher temperatures, the individual $C_q$ values converge more rapidly, as shown Fig.~\ref{fig:average}b. Since NMR experiments operate on the \si{\mega\hertz} regime they average over these slow dynamics. In contrast, our present simulation fails to fully converge when the EFG tensor is averaged over the trajectory of individual nitrogen atoms prior to diagonalisation.

The core issue is that the orientation distributions of some individual molecules ({\em e.g.}, Fig.~\ref{fig:average}d) differ from the fully averaged ensemble distribution (Fig.~\ref{fig:cq}e). However, as discussed previously, the symmetry of the system ---specifically the \SI{180}{\degree} rotational symmetry in the $ab$-plane within each orientation family and the equivalence of both families--- ensures that all nitrogen atoms should contribute statistically equivalent EFG tensors when sampled over a sufficiently long time. Therefore, spatial averaging of the EFG tensor across all nitrogen atoms in the $8\times8\times8$ supercell prior to diagonalisation is not only justified, but statistically superior. It accelerates convergence, suppresses artefacts due to slow molecular reorientation, and yields results in excellent agreement with experiment. As seen from the slow decay in Fig.~\ref{fig:average}a, per-atom EFG contributions have not fully converged even after \SI{2}{\nano\second}, reinforcing the need for averaging of all nitrogen atoms to achieve reliable $C_q$ values. At this point, it is crucial to carefully analyse the underlying physics; a naive computational approach relying on overly short trajectories can lead to non-converged results. 

A similarly careful analysis is required when evaluating the EFG for the iodine atoms. Symmetry suggests the existence of two types of iodine atoms: those within the \ce{PbI2} plane and those within the \ce{MAI} plane (see Fig.~\ref{fig:struc}). As explained earlier and towards the end of the caption of Fig.~\ref{fig:cq}, in the tetragonal phase, the hydrogen atoms in the \ce{NH3} group form hydrogen bonds with the iodine atoms in the \ce{PbI2} plane but not with those in the \ce{MAI} plane. For iodine, it is generally appropriate to average the EFG tensors over the trajectories of individual iodine atoms, diagonalise the time-averaged tensors, and then analyse the results. However, to improve statistical reliability, a better approach is to categorise the iodine atoms into three groups based on their \ce{Pb-I} bond orientation along the lattice directions $a$, $b$, and $c$, respectively. The EFG tensors can then be averaged over all atoms and time steps within each group before diagonalisation, yielding three representative tensors. This is the approach we used to generate Fig.~\ref{fig:cq}d. The two iodine atoms in the \ce{PbI2}-layer yield an identical and smaller $C_q$, whereas the iodine atom in the \ce{MAI}-layer yields the larger $C_q$ value. Alternatively, one could perform simulations of several \SI{100}{\nano\second}, average the EFG tensor over individual atoms, and diagonalise these individual tensors. We are confident that this would yield results identical to the present results, albeit at a much greater computational cost. 

\section{Conclusion}
\label{sec:conclusion}

The present study showcases the effectiveness of a computational framework that combines first-principles simulations with machine-learning methodologies. By employing MLFFs to access previously unattainable timescales, we have developed a robust approach to accurately predict the quadrupolar coupling constant in \ce{MAPbI3} and capture its behaviour across phase transitions. Our method integrates thermostat-driven molecular dynamics sampling with machine-learned EFG models, leveraging three-body descriptors that respect the inherent symmetry of the EFG tensor. This framework successfully reproduces the temperature-dependent evolution of the \ce{^{14}N} and \ce{^{127}I} quadrupolar coupling constant and predicts the tetragonal-to-cubic phase transition temperature with an error of less than \SI{2}{\percent} relative to experimental values.

In contrast to conventional first-principles molecular dynamics, which are computationally prohibitive for long time scales and large system sizes, the approach significantly reduces the computational cost while maintaining high accuracy. The combination of MLFFs and symmetry-preserving EFG descriptors ensures both efficiency and transferability, making the method very convenient for studying temperature-dependent quadrupolar interactions in complex materials. 

Overall, our findings underscore the potential of advanced MLFF techniques to provide deeper insights into the complex phase behaviour of hybrid perovskites. Future work will focus on further refining the MLFF, exploring additional descriptors, and extending the approach to other technologically relevant materials. Our results highlight the power of machine learning in enhancing computational spectroscopy, offering a scalable, accurate, and efficient tool for predicting phase transitions and NMR observables in complex systems.


\section{Acknowledgements}
This research was funded in whole by the Austrian Science Fund (FWF) 10.55776/F8100. For open access purposes, the author has applied a CC BY public copyright license to any author accepted manuscript version arising from this submission. The computational results presented have been achieved in part using the Vienna Scientific Cluster (VSC).



\bibliography{ref.bib}
\end{document}